# High carrier mobility epitaxially aligned PtSe$_2$ films grown by one-zone selenization

*Michaela Sojková$^{1*}$, Edmund Dobročka$^1$, Peter Hutár$^1$, Valéria Tašková$^1$, Lenka Pribusová-Slušná$^1$, Roman Stoklas$^1$, Igor Píš$^2$, Federica Bondino$^2$, Frans Munnik$^3$ and Martin Hulman$^1$*

$^1$Institute of Electrical Engineering, SAS, Dúbravská cesta 9, 841 04 Bratislava, Slovakia

$^2$IOM-CNR, Laboratorio TASC, S.S. 14 km 163.5, 34149 Basovizza, Trieste, Italy

$^3$Helmholtz-ZentrumDresden-Rossendorf, e.V. Bautzner Landstrasse 400, D-01328 Dresden, Germany

**Corresponding author**

Email: michaela.sojkova@savba.sk

**ABSTRACT:** Few-layer $PtSe_2$ films are promising candidates for applications in high-speed electronics, spintronics and photodetectors. Reproducible fabrication of large-area highly crystalline films is, however, still a challenge. Here, we report the fabrication of epitaxially aligned $PtSe_2$ films using one-zone selenization of pre-sputtered platinum layers. We have studied the influence of growth conditions on structural and electrical properties of the films prepared from Pt layers with different initial thickness. The best results were obtained for the $PtSe_2$ layers grown at elevated temperatures (600 °C). The films exhibit signatures for a long-range in-plane ordering resembling an epitaxial growth. The charge carrier mobility determined by Hall-effect measurements is up to 24 $cm^2$/V.s.

1. INTRODUCTION

Transition metal dichalcogenides (TMDCs) – layered materials stacked with a weak out-of-plane van der Waals interaction, belong to the group of two-dimensional materials. TMDCs include over 40 different combinations of transition metal from group IV to VI and chalcogens (S, Se and Te) [1]. Up to now, $MoS_2$ has been the most studied material from this group [2,3]. However, for the high-speed electronic application its low mobility seems to be the limiting factor. Thus new TMDC materials with better properties are desirable [4]. It is known that the charge carrier mobility of the TMDC and 2D semiconducting materials is influenced by several factors, such as charge carrier polarity, temperature, number of layers, substrate, charged impurities, localized states, defects, device geometry, and contacts [5]. This is the reason why significant discrepancies can be found in the electronic transport of mono- and few-layer form.

For example, for a graphene monolayer, the mobility reaches 140,000 cm$^2$/V.s but it does not exceed 10,000 cm$^2$/V.s [5] for the bulk. The carrier mobility of black phosphorous (BP) is much lower than those of graphene (over 1000 cm$^2$/V.s), however, the current on-off ratio (more than 10$^5$) is almost four orders of magnitude higher than that of graphene [6]. After encapsulation in h-BN, the hole mobility of BP can reach up to 45,000 and 5,000 cm$^2$/V.s at 2 K and room temperature, respectively [7]. The single-layer silicene yields a field-effect mobility of ~ 100 cm$^2$/V.s, while multilayer silicene showed similar mobility of about 200 cm$^2$/V.s [8]. The carrier mobility in ML antimonene was estimated to be 150/510 cm$^2$/V.s for electrons/holes [9]. Electrical characterizations of a MoS$_2$ monolayer have shown n-type conductivity with a room temperature mobility up to 200 cm$^2$/V.s [10] which is rather low compared to the mobility (200 − 500 cm$^2$/V.s) of a bulk MoS$_2$ [11]. The high-throughput calculations based on density functional theory (DFT) are used to identify 2D semiconductors with a suitable bandgap and higher mobilities. Zhang et al. [12] have performed the electronic calculations on the selected TMDCs with a general formula MX$_2$ (M = Mo, W, Sn, Hf, Zr or Pt; X = S, Se or Te). They have found several compounds (e.g. MoTe$_2$, HfSe$_2$, ZrSe$_2$, PtSe$_2$) with the photon limited mobility values exceeding 1000 cm$^2$/V.s.

Platinum diselenide (PtSe$_2$) has attracted interest due to its semi-metallic electronic structure, high optoelectronic performance, and enhanced photocatalytic activities [13]. Monolayer (ML) PtSe$_2$ is a semiconductor with a band-gap of ≈ 1.2 eV [14]. High values of predicted charge-carrier mobility together with improved stability compared to other TMDCs [15] makes PtSe$_2$ a promising candidate for high-speed electronics [4]. It has been shown that PtSe$_2$ can find applications also for photocatalytic activity [12,16,17], photo-detection [18] and quick-response gas sensing [18,19]. For all these applications, high quality large-area PtSe$_2$ films are necessary.

Several methods have been reported for the growth of monolayer and few-layer PtSe$_2$ films. Yan et al. [4] presented epitaxial growth of high-quality PtSe$_2$ thin films with controlled thickness using molecular beam epitaxy. Although this method offers the possibility to grow large size single- crystalline films on a variety of substrates, it is less suitable for device fabrication on a large scale. Chemical vapor deposition [20,21] and the selenization of thin platinum films [22–25] are currently widely used methods for the fabrication of PtSe$_2$ layers in a nanometer thickness range. Large-area (~ cm$^2$) films can be prepared by selenization, however, the nanocrystalline character is an inherent property of thin PtSe$_2$ layers grown using this method. The alignment of PtSe$_2$ atomic layers with respect to the substrate can also be controlled [24]. On the other hand, the nanocrystals are oriented randomly and no long-range ordering within the plane of the layer has been achieved experimentally, so far. As a result, the as-measured charge carrier mobility in selenized films is much lower than > 1000 cm$^2$/V.s predicted theoretically [12]. Typical experimental values span a range from below 1 to 10 – 15 cm$^2$/V.s [22,23,25].

In this paper, we present the fabrication of few-layer PtSe$_2$ films by one-zone selenization. The substrate with pre-deposited platinum and the selenium powder are placed next to each other in the center of the furnace. Recently, we have used this method for the preparation of MoS$_2$ thin films grown on different substrates with possibility to control the layer alignment [26–29]. In this paper, we study how the selenization temperature, the growth time and the thickness of an initial Pt layer affect the structural and electrical properties of as-prepared PtSe$_2$ films. For characterization of the structural and chemical properties of thin films different methods, such as X-ray diffraction measurements (XRD), Raman spectroscopy, X-ray photoelectron spectroscopy (XPS) and others have been used. We have found that $PtSe_2$ layers grown at elevated temperatures on the sapphire substrate exhibit signatures for a long-range in-plane ordering which resembles an epitaxial growth. The charge carrier mobility determined by Hall-effect measurements is up to 24 $cm^2$/V.s in these films.

## 2. EXPERIMENTAL SECTION

**$PtSe_2$ fabrication** $PtSe_2$ thin films were prepared by a two-step method on the top of the c-plane (0001) sapphire substrate. The substrate dimensions were $10 \times 10$ $mm^2$. Before the deposition, we annealed the substrate at 1000 °C for 3 h in the air to decrease the stress and number of defects [30]. At first, DC magnetron sputtering in Ar atmosphere ($10^{-3}$ mbar) from a platinum target at room temperature was used for the fabrication of Pt films. The DC power and emission current were set to 580 W and 0.18 A, respectively. The thickness of the as-prepared Pt films was controlled by the rotation speed of a sample holder. Further, the pre-deposited Pt layers were selenized in a custom-designed CVD chamber. In particular, the Pt layer was annealed in selenium vapors at a high temperature of 400 – 700 °C in the $N_2$ atmosphere at ambient pressure. The substrate together with the Se powder were placed in the center of the furnace, so the substrate and the powder temperature were the same during the growth [26,27] contrary to the standard CVD method where the reaction takes place in a two-zone furnace with the selenium powder and the Pt substrate heated at different temperatures.

**Chemical composition analyses**

Raman measurements were performed on a confocal Raman microscope (Alpha 300R, WiTec, Germany) using an excitation laser with a 532 nm wavelength. The laser power was kept as low as 1 mW to avoid any beam damage. The scattered Raman signal was collected by a 50× (NA = 0.8) microscope objective and detected by a Peltier-cooled EMCCD camera. For dispersing the Raman spectra, a blazed grating with 1800 grooves/mm was employed. The spectral resolution of the entire Raman system is about 0.75 cm$^{-1}$. The Raman spectra were acquired at ambient conditions.

Synchrotron radiation photoelectron spectroscopy was performed using the BACH beamline at the Elettra synchrotron in Trieste, Italy.[31,32] The beamline was equipped with a Scienta R3000 hemispherical analyzer[33] at an angle of 60° respective to the X-ray incidence direction. The spectra were recorded in a normal emission geometry, with light linearly polarized in the horizontal plane. The core-level spectra were measured at a photon energy of 702 eV with a total instrumental resolution of 0.2 eV. The binding energy scale was calibrated using the Au $4f_{7/2}$ peak (84.0 eV) of a clean gold reference. Pt 4f and Se 3d core levels were decomposed into their spectral components using Voigt line shapes and Shirley-type background. To calculate the atomic concentrations, the areas of photoemission peaks were corrected for the photon flux, analyzer transmission function, inelastic mean free paths, photoionization cross-section and asymmetry factors [34,35]. The samples were degassed in a vacuum at a temperature of 250 °C for 10 minutes before the measurements.

Rutherford Backscattering Spectrometry (RBS) was used to analyze the sample composition using a 1.7 MeV He$^+$ beam. Backscattered ions have been detected at an angle of 140° to the beam direction, while the angle between sample normal and incoming beam was 18°. The measurements were analyzed with WiNDF V9.6i software.[36]

**Structural analyses** The structural analysis of PtSe$_2$ layers was performed with a diffractometer Bruker D8 DISCOVER equipped with a rotating anode (Cu-Kα) working at the power of 12kW.

The crystal structure of the thin films was studied by X-ray diffraction (XRD) in a symmetrical θ/2θ configuration. The thickness of PtSe$_2$ layers was determined by X-ray reflectivity (XRR). The crystallographic orientation and the texture of the films were studied by the azimuthal (φ-scan) and rocking curve (ω-scan) measurements, respectively.

**Surface analysis** The atomic force microscopy (AFM) measurements were performed in a tapping mode (Bruker, Dimension Edge), using an etched silicon probe (Bruker, RTESPA - 300). The root-mean-square (RMS) surface roughness of the films was calculated using QWYDDION software applying the formula:

$$RMS = \sqrt{\sum(z_i - z_{ave})^2/N},$$

where $z_{ave}$ is the average of the z values within the given area, $z_i$ is the N value for a given point, and ) is the number of measured points within the given area [37].

Scanning electron microscopy (SEM) was also used to investigate the surface morphology of the films (FEG 250).

**Electrical characterization** The Hall coefficient and the resistivity were measured by the Van der Pauw (VdP) method. The charge carrier mobility was then calculated from these quantities. The measurement configuration consists of: i) an electromagnet to set the magnetic field up to 1T by a high power source, ii) a current source Keithley 2400 to set and hold the current on a constant value, and iii) a multimeter Keithley 2700 to measure the resistance and Hall voltage. Indium contacts were used.

## 3. RESULTS AND DISCUSSION

Few-layer PtSe$_2$ films were synthesized by one-zone selenization of the pre-sputtered Pt layers (0.5, 1 and 3 nm thick) on top of a c-plane (0001) sapphire substrate under different growth conditions. The temperature varied from 400 to 700 °C. At 400 °C, the films begin to grow as early as 5 minutes after reaching the maximum temperature. However, better film crystallinity was obtained using a longer annealing time.

Figure 1a and 1b shows Raman spectra of PtSe$_2$ films selenized for 120 minutes at 400 °C and 600 °C, respectively revealing lines belonging to 1T PtSe$_2$: the E$_g$ mode at around 175 cm$^{-1}$ and the A$_{1g}$ mode at ~ 207 cm$^{-1}$. The intensity in the spectra was normalized to the height of the A$_{1g}$ line. As expected, the ratio A$_{1g}$ / E$_g$ increases as the number of PtSe$_2$ layers increases [38]. Besides the E$_g$ and A$_{1g}$ peaks, also a peak at ∼240 cm$^{-1}$ was observed in all samples, being the most intense one in the case of the thinnest films (0.5 nm Pt). This peak was attributed to a longitudinal optical (LO) mode. The latter is a combination of the out-of-plane (A$_{2u}$) and in-plane (E$_u$) vibrations of platinum and selenium atoms, respectively [4]. The LO mode splits into two peaks for the films prepared from 0.5 nm Pt layer and the splitting is more pronounced in the case of the samples selenized at higher temperature. A similar splitting of the LO mode was observed for PtSe$_2$ monolayer films prepared by molecular beam epitaxy [4].

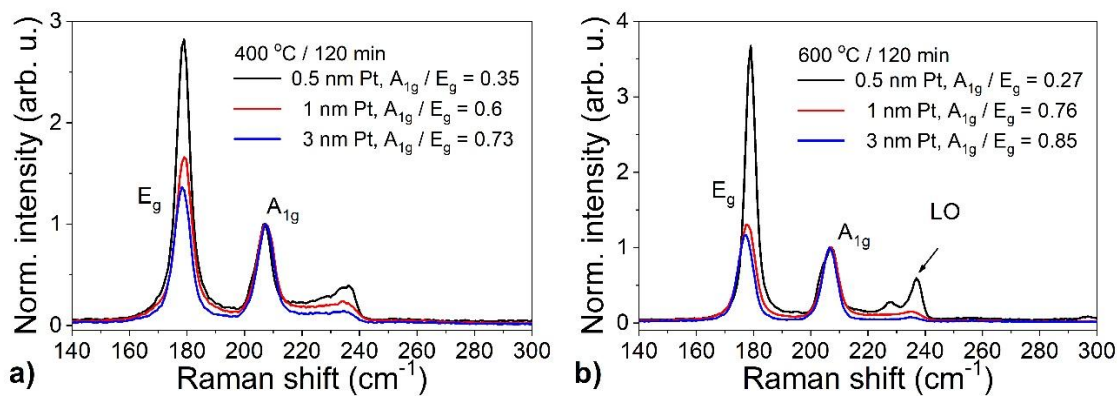

**Figure 1.** (a) Normalized Raman spectra of PtSe$_2$ films prepared at (a) 400 °C for 120 min and (b) at 600 °C for 120 min.

We have increased the selenization temperature up to 700 °C. However, such a high temperature led to a deterioration in the quality of PtSe$_2$ films (see Figure S1 in Supplementary material).

The chemical composition of the PtSe$_2$ films was probed by synchrotron-radiation high-resolution XPS at 702 eV photon energy. Only the presence of platinum and selenium was observed in a wide survey scan. Figure 2 shows Pt 4f and Se 3d spectra of the films prepared at 400 °C and 600 °C / 120 min. Two components can be distinguished in the spectra. The binding energies of the primary peaks, 73.01 ± 0.06 eV for Pt 4f$_{7/2}$ and 54.36 ± 0.07 eV for Se 3d$_{5/2}$, are in a good agreement with the reported values for PtSe$_2$.[14,20,38] The Se$_{3d}$/Pt$_{4f}$ integrated intensity ratio for the sample annealed at 400°C and 600°C is 0.49±0.01 and 0.51±0.01, respectively. After taking into account the atomic sensitivity factors, the intensities yield the Se/Pt atomic ratio of ~ 2, as expected for PtSe$_2$. However, the difference in the ratios indicates a slightly lower relative concentration of Se in the sample treated at 400°C. The secondary XPS components suggest the presence of a second phase with a Se3d$_{5/2}$ peak at ~54.7 eV and Pt 4f$_{7/2}$ at ~71.8 eV. We exclude the possibility of separated Pt and Se phases, because in that case an unreacted metallic Pt would have the Pt 4f component at 71.0 eV. Therefore, we ascribe the minor XPS features to the $PtSe_2$-based phase with under-stoichiometric Se content. XPS spectra taken at higher photon energies (i.e., with larger information depth) showed that the relative amount of the second phase is lower in bulk. Nevertheless, higher intensity of the $PtSe_2$ components in XPS spectra demonstrated that the films prepared at 600 °C were of better chemical purity than those prepared at 400 °C.

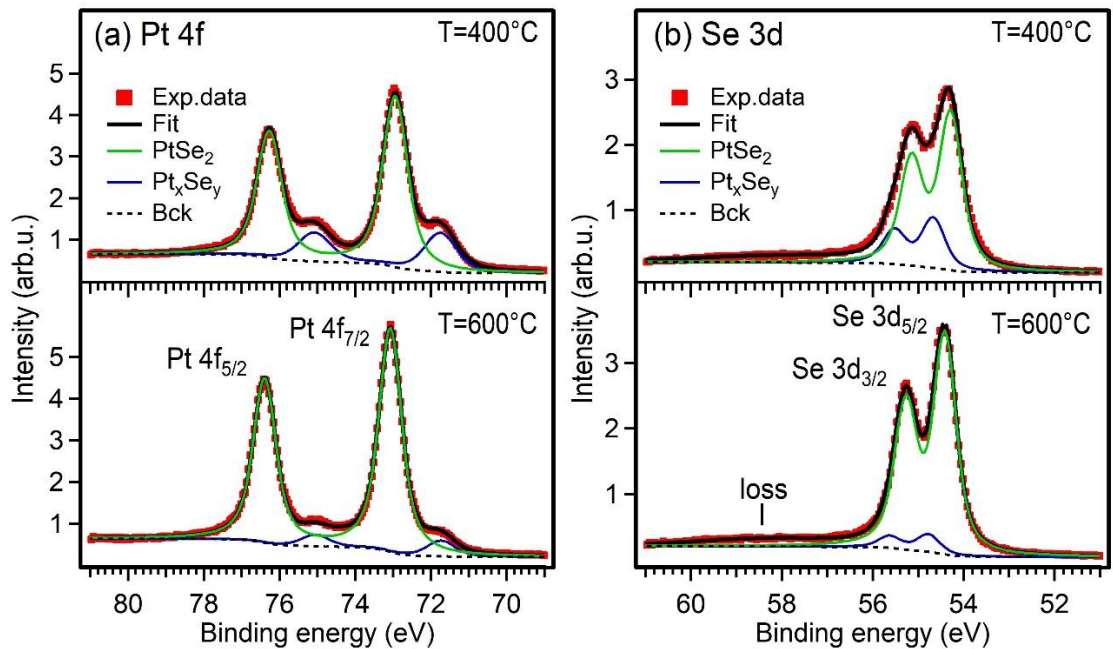

**Figure 2.** High-resolution XPS spectra of $PtSe_2$ films prepared from 3 nm thick layer annealed at 400 and 600 °C for 120 min.

The composition of the films was analyzed using Rutherford Backscattering Spectrometry (RBS). Figure 3 shows results for the $PtSe_2$ thin films grown on the c-plane sapphire substrate at 400 and 600 °C, respectively. Two peaks observed in each spectrum belong to selenium and platinum. The steep edges at lower energy (channels) are due to the substrate. The fit (black line) is in excellent agreement with the measurement. The concentrations of both elements, selenium and platinum in the layer were estimated from the fit. The Se/Pt ratios are 2.05±0.03 for films prepared at 400 °C and 1.92±0.03 for those fabricated at 600 °C, in agreement with the results obtained from synchrotron radiation high-resolution XPS measurements.

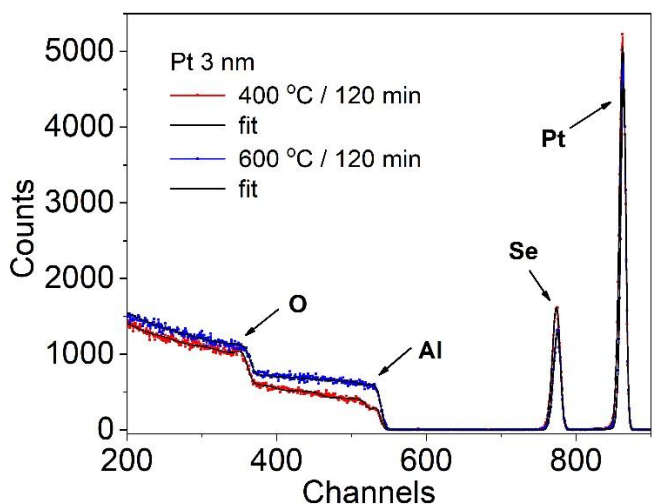

**Figure 3.** RBS spectra of PtSe$_2$ films prepared from 3 nm thick layer annealed at 400 and 600 °C for 120 min.

The structure of as-prepared films was examined by XRD measurements. Figure 4 shows XRD patterns of PtSe$_2$ thin grown at 400 and 600 °C for 120 minutes from 1 and 3 nm thick Pt layers measured in a symmetrical θ/2θ configuration. Only a 0001 diffraction is visible in the patterns suggesting a strong texturation with the c-axis of the PtSe$_2$ layer perpendicular to the substrate. A rise in the diffraction intensity with increased thickness and selenization temperature is related to the total volume of the PtSe$_2$ phase in the sample. A slight up-shift of the diffraction peaks indicates a smaller inter-layer distance in the thicker samples. However, it was not possible to quantify the lattice parameter change with sufficient precision because the measurements were not performed in a high-resolution mode and a single peak was observed in the diffraction pattern. Comparing the diffraction results of the films grown at a higher temperature to those prepared at 400 °C, it is clear that higher temperature increases the crystallinity of the films. It is, in particular, evident from the Laue oscillations appeared on both sides of the dominant 0001 diffraction. The oscillations are interferences due to a finite thickness of the film and are signatures of coherence through the film, thickness uniformity and a negligible roughness of the film surface [39]. The position of satellite maximums and the width of the central peak depend on the thickness of the thin film. This thickness can be determined from the positions of adjacent satellite maxima $\theta_{i+1}$ and $\theta_i$ according to $d = \lambda / (2(\sin\theta_{i+1} - \sin\theta_i))$, where $\lambda$ is the wavelength of the X-ray radiation [39]. For selenization at 600 °C / 120 min, the calculated thickness was 4.5 and 14 nm for PtSe$_2$ grown from 1 and 3 nm thick Pt layer, respectively.

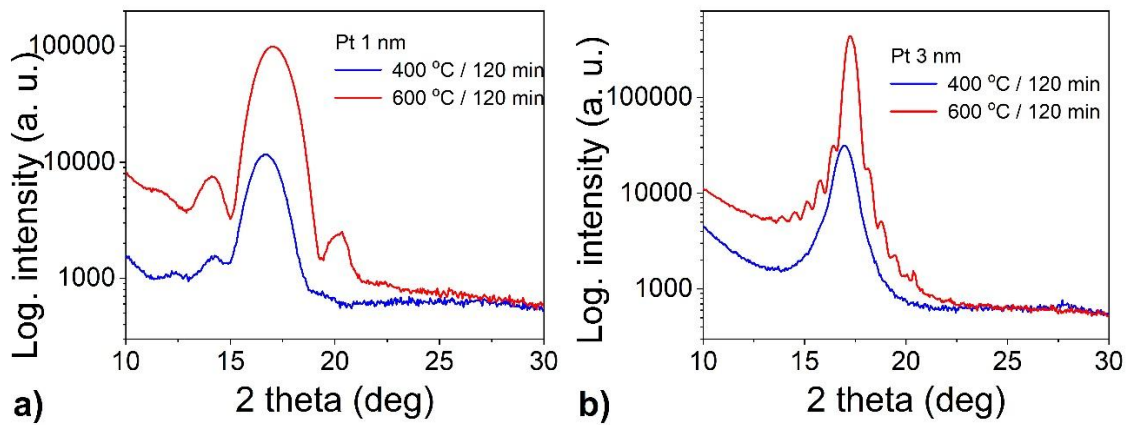

**Figure 4.** XRD pattern of PtSe$_2$ films prepared from (a) 1 and (b) 3 nm thick Pt at 400 and 600 °C for 120 min.

The preferred orientation of the PtSe$_2$ layers was evaluated from $\omega$-scans of the 0001 diffraction peak. The maxima of $\omega$-scans are rather narrow, FWHM (full-width at half maximum) of the curves is in the order of ~ 0.1°, indicating a high degree of texture in the layers. However, in most cases, the $\omega$-scans comprise two maxima with varying angular separation. We can conclude that the PtSe$_2$ layers comprise of two slightly misoriented "sublayers" – one perfectly matched to the surface plane and one declined in the direction of substrate miscut. For a detailed description of the measurement setup and results see Supplementary material (Figure S2, S3).

The in-plane ordering of the PtSe$_2$ layers was determined from $\varphi$-scans. We have chosen the strongest $10\bar{1}1$ diffraction of the hexagonal PtSe$_2$ phase for analysis. The results are given in Figure 5a. As seen, a longer selenization time and especially higher growth temperature lead to higher peak intensity and, therefore, to better crystallinity of the films. The presence of distinct maxima in $\varphi$-scans indicates a tendency of the layers to grow epitaxially. On the other hand, the maxima are rather broad suggesting that an in-plane disorder is still present. We suppose that the peak broadening was caused by the weak forces (Van der Waals) between the layers in the c-direction. To determine the orientation relationship between the layer and the substrate, $\varphi$-scan of the diffraction $10\bar{1}4$ of the sapphire was measured. Following the trigonal symmetry of the sapphire, three extremely narrow maxima (black curve) can be recognized in Figure 5a. Based on the mutual position of the maxima in the $\varphi$-scans of the PtSe$_2$ layer and the sapphire, one can conclude that the hexagonal lattice of the layer is rotated by 30° with respect to the substrate lattice. The possible coincidence of both lattices is schematically shown in Figure 5b. Crystallographic orientation relationship can be expressed as

$PtSe_2(0001)[10\bar{1}0] \| sapphire(0001)[2\bar{1}\bar{1}0]$.

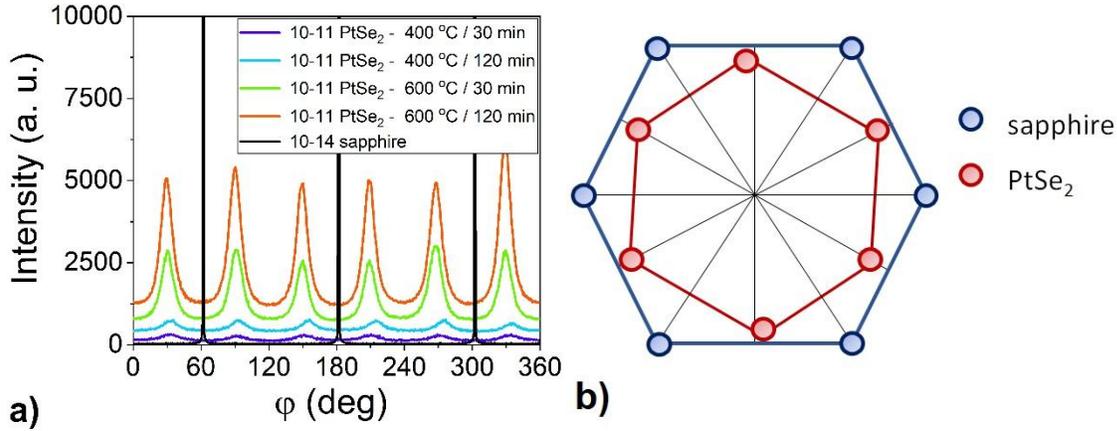

**Figure 5.** (a) $\varphi$-scan of 101 diffraction of PtSe$_2$ films prepared from 3 nm thick Pt at 400 and 600 °C for 30 or 120 min. (b) Schematic growth of PtSe$_2$ films on the c-plane sapphire substrate.

The rotation of the layer lattice helps to overcome the significant difference of more than 20% between the in-plane lattice parameters of the sapphire substrate and the PtSe$_2$ layer.

As-prepared PtSe$_2$ films continuously cover the substrate (See SEM images in the Supplementary material, Figure S4, S5). The surface of the films is very smooth as confirmed by AFM. The surface roughness represented by an RMS (root mean square) value is in the range from 0.4 to 1 nm for all the PtSe$_2$ films grown at 400 and 600 °C (see Supplementary, Figure S6).

The low roughness and high crystallinity of the films allowed us to measure the X-ray reflectivity (XRR) and to determine the thickness of the PtSe$_2$ (Figure 6). The intensity oscillations observed in the XRR spectrum are so-called Kiessig fringes due to the interference of the x-ray beams reflected from the upper and bottom interfaces of the PtSe$_2$ film. The parallelism of the interfaces and low surface roughness are necessary conditions for the fringes to occur. The distance between the fringe maxima is inversely proportional to the film thickness [39]. The latter was estimated from the fits to the XRR spectra shown in Figure 6.

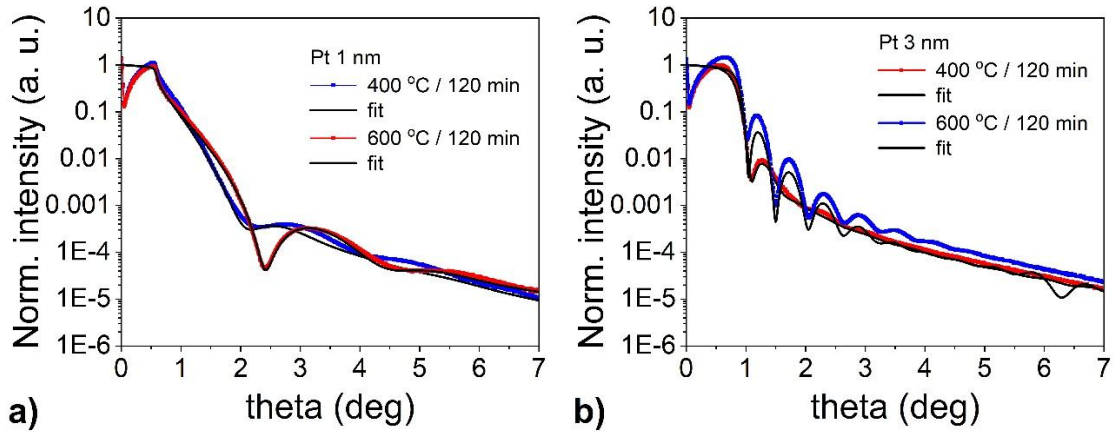

**Figure 6.** XRR pattern of PtSe$_2$ films prepared from 1 and 3 nm thick Pt at (a) 400 °C and (b) 600 °C for 120 min.

Table 1. presents the thickness values calculated from XRR fits and Laue oscillations for PtSe$_2$ films prepared under diverse growth conditions. They show a very good agreement with each other. The thickness of the PtSe$_2$ layers is larger by a factor of ~ 5 than the thickness of the Pt films from which the layers grew.

| Initial Pt thickness | Selenization conditions | Thickness calculated from Laue oscillations | Thickness calculated from XRR |
|---|---|---|---|
| 1 nm | 400 °C / 30 min | 5.0 nm | 5.2 nm |
| 3 nm | | 13.2 ± 0.3 nm | 13.8 nm |
| | | | |
| 1 nm | 400 °C / 120 min | 4.8 nm | 4.7 nm |
| 3 nm | | 12.1 nm | 12.1 nm |
| | | | |
| 1 nm | 600 °C / 30 min | 3.9 ± 0.2 nm | 4.4 nm |
| 3 nm | | 12.5 ± 0.3 nm | 13.1 nm |
| | | | |
| 0.5 nm | 600 °C / 120 min | 3.1 nm | 2.8 nm |
| 1 nm | | 4.5 ± 0.2 nm | 4.5 nm |
| 3 nm | | 14.0 ± 0.5 nm | 14.3 nm |

**Table 1.** The thickness of PtSe$_2$ films prepared under different conditions as estimated from Laue oscillations and Kiessig fringes.

We performed Hall effect measurements in Van der Pauw configuration to determine the charge carrier mobility in the PtSe$_2$ films. Table 2. shows the results for the PtSe$_2$ films with 0.5 – 5 nm initial Pt thickness fabricated under different conditions.

| Initial Pt thickness | Selenization conditions | μ (cm$^2$/Vs) |
|---|---|---|
| 1 nm | 400 °C / 30 min | 5.1 |
| 3 nm |  | 4.42 |
|  |  |  |
| 1 nm | 400 °C / 120 min | 5.63 |
| 3 nm |  | 7.81 |
|  |  |  |
| 0.5 nm | 600 °C / 30 min | 23.74 |
| 1 nm |  | 12.75 |
| 3 nm |  | 14.86 |
| 5 nm |  | 7.62 |
|  |  |  |
| 0.5 nm | 600 °C / 120 min | 8.53 |
| 1 nm |  | 18.53 |
| 3 nm |  | 2.95 |

**Table 2.** The charge carrier mobility of PtSe$_2$ films with different thickness prepared under different conditions.

Some conclusions can be drawn from the values shown in Table 2. In general, the samples prepared at higher temperatures show better mobility than those grown at 400 °C for the same time. However, the relation of other parameters to the mobility values is harder to recognize. In particular, we do not observe any conclusive mobility dependence on film thickness. Also, a prolonged time of selenization does not lead to higher mobility. The layer grown from a 0.5 nm thick Pt film and selenized at 600 °C for 30 min has the highest mobility of 23.7 cm$^2$/V.s among all the samples. Our values are higher than the Hall mobilities observed in selenized samples [22,25] and field-effect mobilities [20,21]. Yim et al. [22] reported Hall hole mobility of 3.5 cm$^2$/V.s on the PtSe$_2$ synthesized from the 1 nm thick Pt initial layer. Ansari et al.

[25] published the fabrication of PtSe$_2$ films made from 0.7, 1 and 1.5 nm thick Pt films with the Hall hole mobility values in the range of 5 to 13 cm$^2$/V.s.

## 4. CONCLUSION

We have successfully fabricated large-area PtSe$_2$ few-layer films on the c-plane sapphire substrate by one-zone selenization of pre-deposited Pt layers. The structure and chemical composition of PtSe$_2$ samples were investigated by several methods – Raman spectroscopy, synchrotron-radiation high-resolution XPS, RBS. The results obtained from XRD and XRR measurements confirm the high quality of the films. PtSe$_2$ layers are aligned horizontally with the c-axis perpendicular to the substrate. The layers have a well-defined crystallographic relationship to the substrate and a long-range in-plane ordering that resembles an epitaxial growth. The presence of Laue oscillations and Kiessig fringes in XRD and XRR measurements, respectively, allowed us to calculate the film thickness. It is larger by a factor of 5 than that of the platinum before selenization. The charge carrier mobility of the films determined by Hall-effect measurements is up to 24 cm$^2$/V.s. We suppose that high crystalline ordering together with low roughness and good chemical purity ensured the charge carrier mobility values higher than those from previous reports.

**Author Contribution:**

MS performed the experimental syntheses and characterizations of $PtSe_2$ layers; ED performed the XRD measurements and interpretation of the results; PH, VT and LPS performed the Raman characterization; IP and FB performed the XPS measurements and interpretation of the results; FM performed the RBS measurements and interpretation of the results; MS and MH supervised the project and wrote the manuscript. All authors discussed the results and commented on the manuscript.

**Declaration of Competing Interests**

The authors declare that they have no known competing financial interests or personal relationships that could have appeared to influence the work reported in this paper.

**ACKNOWLEDGEMENTS**

This project has received funding from the EU-H2020 research and innovation programme under grant agreement No. 654360 having benefitted from the access provided by IOM-CNR in Trieste (Italy) within the framework of the NFFA-Europe Transnational Access Activity. This study was performed during the implementation of the project Building-up Centre for advanced materials application of the Slovak Academy of Sciences, ITMS project code 313021T081 supported by Research & Innovation Operational Programme funded by the ERDF. This work was supported by the Slovak Research and Development Agency, APVV-15-0693, APVV-17-0352, APVV-17-0560, APVV-19-0365 and Slovak Grant Agency for Science, VEGA 2/0149/17. We acknowledge Ján Dérer for the deposition of platinum films. Parts of this research were carried out at the IBC at the Helmholtz-Zentrum Dresden-Rossendorf e. V., a member of the Helmholtz Association.

**Supplementary material**

**Supporting Information:** Additional characterization data including detailed explanation of ω-scans, SEM and AFM images.

**REFERENCES**

[1] M. Chhowalla, H.S. Shin, G. Eda, L.-J. Li, K.P. Loh, H. Zhang, The chemistry of two-dimensional layered transition metal dichalcogenide nanosheets, Nature Chemistry. 5 (2013) 263–275. https://doi.org/10.1038/nchem.1589.

[2] U. Krishnan, M. Kaur, K. Singh, M. Kumar, A. Kumar, A synoptic review of MoS2: Synthesis to applications, Superlattices and Microstructures. 128 (2019) 274–297. https://doi.org/10.1016/j.spmi.2019.02.005.

[3] H. Wang, C. Li, P. Fang, Z. Zhang, J.Z. Zhang, Synthesis, properties, and optoelectronic applications of two-dimensional $MoS_2$ and $MoS_2$-based heterostructures, Chemical Society Reviews. 47 (2018) 6101–6127. https://doi.org/10.1039/C8CS00314A.

[4] M. Yan, E. Wang, X. Zhou, G. Zhang, H. Zhang, K. Zhang, W. Yao, N. Lu, S. Yang, S. Wu, T. Yoshikawa, K. Miyamoto, T. Okuda, Y. Wu, P. Yu, W. Duan, S. Zhou, High quality atomically thin $PtSe_2$ films grown by molecular beam epitaxy, 2D Materials. 4 (2017) 045015. https://doi.org/10.1088/2053-1583/aa8919.

[5] M. Velický, P.S. Toth, From two-dimensional materials to their heterostructures: An electrochemist's perspective, Applied Materials Today. 8 (2017) 68–103. https://doi.org/10.1016/j.apmt.2017.05.003.

[6] Y. Xu, Z. Shi, X. Shi, K. Zhang, H. Zhang, Recent progress in black phosphorus and black-phosphorus-analogue materials: properties, synthesis and applications, Nanoscale. 11 (2019) 14491–14527. https://doi.org/10.1039/C9NR04348A.

[7] G. Long, D. Maryenko, J. Shen, S. Xu, J. Hou, Z. Wu, W.K. Wong, T. Han, J. Lin, Y. Cai, R. Lortz, N. Wang, Achieving Ultrahigh Carrier Mobility in Two-Dimensional Hole Gas of Black Phosphorus, Nano Letters. 16 (2016) 7768–7773. https://doi.org/10.1021/acs.nanolett.6b03951.

[8] A. Molle, C. Grazianetti, L. Tao, D. Taneja, Md.H. Alam, D. Akinwande, Silicene, silicene derivatives, and their device applications, Chemical Society Reviews. 47 (2018) 6370–6387. https://doi.org/10.1039/C8CS00338F.

[9] Y. Wang, P. Huang, M. Ye, R. Quhe, Y. Pan, H. Zhang, H. Zhong, J. Shi, J. Lu, Many-body Effect, Carrier Mobility, and Device Performance of Hexagonal Arsenene and Antimonene, Chemistry of Materials. 29 (2017) 2191–2201. https://doi.org/10.1021/acs.chemmater.6b04909.

[10] B. Radisavljevic, A. Radenovic, J. Brivio, V. Giacometti, A. Kis, Single-layer MoS2 transistors, Nature Nanotechnology. 6 (2011) 147–150. https://doi.org/10.1038/nnano.2010.279.

[11] R. Fivaz, E. Mooser, Mobility of Charge Carriers in Semiconducting Layer Structures, Physical Review. 163 (1967) 743–755. https://doi.org/10.1103/PhysRev.163.743.

[12] W. Zhang, Z. Huang, W. Zhang, Y. Li, Two-dimensional semiconductors with possible high room temperature mobility, Nano Research. 7 (2014) 1731–1737. https://doi.org/10.1007/s12274-014-0532-x.

[13] A. Kandemir, B. Akbali, Z. Kahraman, S.V. Badalov, M. Ozcan, F. Iyikanat, H. Sahin, Structural, electronic and phononic properties of $PtSe_2$: from monolayer to bulk, Semiconductor Science and Technology. 33 (2018) 085002. https://doi.org/10.1088/1361-6641/aacba2.

[14] Y. Wang, L. Li, W. Yao, S. Song, J.T. Sun, J. Pan, X. Ren, C. Li, E. Okunishi, Y.-Q. Wang, E. Wang, Y. Shao, Y.Y. Zhang, H. Yang, E.F. Schwier, H. Iwasawa, K. Shimada, M. Taniguchi, Z. Cheng, S. Zhou, S. Du, S.J. Pennycook, S.T. Pantelides, H.-J. Gao, Monolayer $PtSe_2$, a New Semiconducting Transition-Metal-Dichalcogenide, Epitaxially Grown by Direct Selenization of Pt, Nano Letters. 15 (2015) 4013–4018. https://doi.org/10.1021/acs.nanolett.5b00964.

[15] Y. Zhao, J. Qiao, Z. Yu, P. Yu, K. Xu, S.P. Lau, W. Zhou, Z. Liu, X. Wang, W. Ji, Y. Chai, High-Electron-Mobility and Air-Stable 2D Layered $PtSe_2$ FETs, Advanced Materials. 29 (2017) 1604230. https://doi.org/10.1002/adma.201604230.

[16] D. Voiry, J. Yang, M. Chhowalla, Recent Strategies for Improving the Catalytic Activity of 2D TMD Nanosheets Toward the Hydrogen Evolution Reaction, Advanced Materials. 28 (2016) 6197–6206. https://doi.org/10.1002/adma.201505597.

[17] S. Ye, W.-C. Oh, Demonstration of enhanced the photocatalytic effect with PtSe2 and TiO2 treated large area graphene obtained by CVD method, Materials Science in Semiconductor Processing. 48 (2016) 106–114. https://doi.org/10.1016/j.mssp.2016.03.001.

[18] C. Yim, K. Lee, N. McEvoy, M. O'Brien, S. Riazimehr, N.C. Berner, C.P. Cullen, J. Kotakoski, J.C. Meyer, M.C. Lemme, G.S. Duesberg, High-Performance Hybrid Electronic Devices from Layered $PtSe_2$ Films Grown at Low Temperature, ACS Nano. 10 (2016) 9550–9558. https://doi.org/10.1021/acsnano.6b04898.

[19] M. Sajjad, E. Montes, N. Singh, U. Schwingenschlögl, Superior Gas Sensing Properties of Monolayer $PtSe_2$, Advanced Materials Interfaces. 4 (2017) 1600911. https://doi.org/10.1002/admi.201600911.

[20] Z. Wang, Q. Li, F. Besenbacher, M. Dong, Facile Synthesis of Single Crystal $PtSe_2$ Nanosheets for Nanoscale Electronics, Advanced Materials. 28 (2016) 10224–10229. https://doi.org/10.1002/adma.201602889.

[21] H. Xu, H. Zhang, Y. Liu, S. Zhang, Y. Sun, Z. Guo, Y. Sheng, X. Wang, C. Luo, X. Wu, J. Wang, W. Hu, Z. Xu, Q. Sun, P. Zhou, J. Shi, Z. Sun, D.W. Zhang, W. Bao, Controlled Doping of Wafer-Scale $PtSe_2$ Films for Device Application, Advanced Functional Materials. 29 (2019) 1805614. https://doi.org/10.1002/adfm.201805614.

[22] C. Yim, N. McEvoy, S. Riazimehr, D.S. Schneider, F. Gity, S. Monaghan, P.K. Hurley, M.C. Lemme, G.S. Duesberg, Wide Spectral Photoresponse of Layered Platinum Diselenide-Based Photodiodes, Nano Letters. 18 (2018) 1794–1800. https://doi.org/10.1021/acs.nanolett.7b05000.

[23] T.-Y. Su, H. Medina, Y.-Z. Chen, S.-W. Wang, S.-S. Lee, Y.-C. Shih, C.-W. Chen, H.-C. Kuo, F.-C. Chuang, Y.-L. Chueh, Phase-Engineered $PtSe_2$-Layered Films by a Plasma-Assisted Selenization Process toward All $PtSe_2$-Based Field Effect Transistor to Highly Sensitive, Flexible, and Wide-Spectrum Photoresponse Photodetectors, Small. 14 (2018) 1800032. https://doi.org/10.1002/smll.201800032.

[24] S.S. Han, J.H. Kim, C. Noh, J.H. Kim, E. Ji, J. Kwon, S.M. Yu, T.-J. Ko, E. Okogbue, K.H. Oh, H.-S. Chung, Y. Jung, G.-H. Lee, Y. Jung, Horizontal-to-Vertical Transition of 2D Layer Orientation in Low-Temperature Chemical Vapor Deposition-Grown PtSe$_2$ and Its Influences on Electrical Properties and Device Applications, ACS Applied Materials & Interfaces. 11 (2019) 13598–13607. https://doi.org/10.1021/acsami.9b01078.

[25] L. Ansari, S. Monaghan, N. McEvoy, C.Ó. Coileáin, C.P. Cullen, J. Lin, R. Siris, T. Stimpel-Lindner, K.F. Burke, G. Mirabelli, R. Duffy, E. Caruso, R.E. Nagle, G.S. Duesberg, P.K. Hurley, F. Gity, Quantum confinement-induced semimetal-to-semiconductor evolution in large-area ultra-thin PtSe2 films grown at 400 °C, Npj 2D Materials and Applications. 3 (2019). https://doi.org/10.1038/s41699-019-0116-4.

[26] M. Sojková, P. Siffalovic, O. Babchenko, G. Vanko, E. Dobročka, J. Hagara, N. Mrkyvkova, E. Majková, T. Ižák, A. Kromka, M. Hulman, Carbide-free one-zone sulfurization method grows thin MoS2 layers on polycrystalline CVD diamond, Scientific Reports. 9 (2019). https://doi.org/10.1038/s41598-018-38472-9.

[27] M. Sojková, K. Vegso, N. Mrkyvkova, J. Hagara, P. Hutár, A. Rosová, M. Čaplovičová, U. Ludacka, V. Skákalová, E. Majková, P. Siffalovic, M. Hulman, Tuning the orientation of few-layer MoS$_2$ films using one-zone sulfurization, RSC Advances. 9 (2019) 29645–29651. https://doi.org/10.1039/C9RA06770A.

[28] M. Hulman, M. Sojková, K. Végsö, N. Mrkyvkova, J. Hagara, P. Hutár, P. Kotrusz, J. Hudec, K. Tokár, E. Majkova, P. Siffalovic, Polarized Raman Reveals Alignment of Few-Layer MoS$_2$ Films, The Journal of Physical Chemistry C. (2019). https://doi.org/10.1021/acs.jpcc.9b08708.

[29] J. Hagara, N. Mrkyvkova, P. Nádaždy, M. Hodas, M. Bodík, M. Jergel, E. Majková, K. Tokár, P. Hutár, M. Sojková, A. Chumakov, O. Konovalov, P. Pandit, S. Roth, A. Hinderhofer, M. Hulman, P. Siffalovic, F. Schreiber, Reorientation of π-conjugated molecules on few-layer MoS$_2$ films, Physical Chemistry Chemical Physics. 22 (2020) 3097–3104. https://doi.org/10.1039/C9CP05728E.

[30] Š. Chromik, M. Cannaerts, Š. Gaži, C. Van Haesendonck, M. Španková, P. Kúš, Š. Beňačka, The influence of (102) sapphire substrate on structural perfection of CeO2 thin films, Physica C: Superconductivity. 371 (2002) 301–308. https://doi.org/10.1016/S0921-4534(01)01098-X.

[31] M. Zangrando, M. Zacchigna, M. Finazzi, D. Cocco, R. Rochow, F. Parmigiani, Polarized high-brilliance and high-resolution soft x-ray source at ELETTRA: The performance of beamline BACH, Review of Scientific Instruments. 75 (2004) 31–36. https://doi.org/10.1063/1.1634355.

[32] M. Zangrando, M. Finazzi, G. Paolucci, G. Comelli, B. Diviacco, R.P. Walker, D. Cocco, F. Parmigiani, BACH, the beamline for advanced dichroic and scattering experiments at ELETTRA, Review of Scientific Instruments. 72 (2001) 1313. https://doi.org/10.1063/1.1334626.

[33] G. Drera, G. Salvinelli, J. Åhlund, P.G. Karlsson, B. Wannberg, E. Magnano, S. Nappini, L. Sangaletti, Transmission function calibration of an angular resolved analyzer for X-ray photoemission spectroscopy: Theory vs experiment, Journal of Electron Spectroscopy and Related Phenomena. 195 (2014) 109–116. https://doi.org/10.1016/j.elspec.2014.06.010.

[34] D.E. Parry, Atomic calculation of photoionization cross-sections and asymmetry parameters J.-J. YEH, Published by Gordon and Breach, Langhorne PA, 1993 ISBN 2-88124-585-4,

Rapid Communications in Mass Spectrometry. 8 (1994) 579–579. https://doi.org/10.1002/rcm.1290080716.

[35] J.J. Yeh, I. Lindau, Atomic subshell photoionization cross sections and asymmetry parameters: $1 \leqslant Z \leqslant 103$, Atomic Data and Nuclear Data Tables. 32 (1985) 1–155. https://doi.org/10.1016/0092-640X(85)90016-6.

[36] N.P. Barradas, C. Jeynes, R.P. Webb, Simulated annealing analysis of Rutherford backscattering data, Applied Physics Letters. 71 (1997) 291–293. https://doi.org/10.1063/1.119524.

[37] W. Wang, Y. Zheng, X. Li, Y. Li, L. Huang, G. Li, High-performance nonpolar *a*-plane GaN-based metal–semiconductor–metal UV photo-detectors fabricated on $LaAlO_3$ substrates, Journal of Materials Chemistry C. 6 (2018) 3417–3426. https://doi.org/10.1039/C7TC05534J.

[38] M. O'Brien, N. McEvoy, C. Motta, J.-Y. Zheng, N.C. Berner, J. Kotakoski, K. Elibol, T.J. Pennycook, J.C. Meyer, C. Yim, M. Abid, T. Hallam, J.F. Donegan, S. Sanvito, G.S. Duesberg, Raman characterization of platinum diselenide thin films, 2D Materials. 3 (2016) 021004. https://doi.org/10.1088/2053-1583/3/2/021004.

[39] M. Birkholz, P.F. Fewster, C. Genzel, Thin film analysis by X-ray scattering, Wiley-VCH, Weinheim, 2006.

# Supporting information

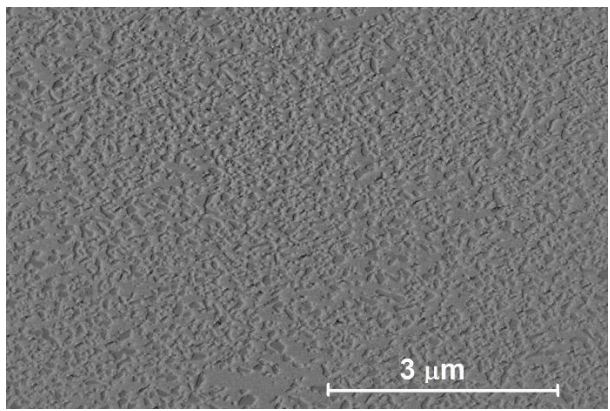

**Figure S1.** SEM image of the 1 nm thick Pt film after selenization at 700 °C for 30 min. Samples selenized at temperatures above 600 °C showed no presence of PtSe$_2$ in Raman measurements. SEM images showed that the film was not continuous and shredded.

Off-set scans $2\theta/\omega$ across both maxima confirmed that they arise from the (0001) lattice planes of PtSe$_2$. In order to elucidate this strange behaviour a series of $\omega$-scans were recorded for seven different values of azimuthal angle $\varphi$ ranging from 0° to 180° with the step size of 30°. Before the measurement the sample was carefully aligned so as the sample surface normal perfectly coincided with the $\varphi$ axis of the goniometer. The result is given in Fig. S2a. It can be seen that the position of the maximum at ~ 8.5° remains unchanged while the position of the second maximum changes with the azimuthal angle $\varphi$. The first maximum can be therefore ascribed to the lattice planes (0001) that are strictly parallel with the sample surface. At the same time, the moving one is associated with the PtSe$_2$ domains slightly declined with respect to the sample surface normal. With changing azimuthal angle $\varphi$ the normal vector (parallel with the corresponding diffraction vector) of the declined domain planes make a precession around the surface normal as is schematically shown in Fig. S2b. In Fig. S2c schematic view in the direction against the surface normal vector is shown. The coloured circles mark the positions of the precessing diffraction vector 0001 around the surface normal and the colours correspond to the particular $\omega$-scans in Fig. S2a. The horizontal line is the projection of the diffraction plane. It has to be pointed out that the $\omega$-scan of a particular diffraction can be measured despite the slight misalignment (up to $\sim \pm 2°$ for standard set-up) of the diffraction vector relative to the diffraction plane. This is a consequence of weak sensitivity of standard diffractometer measurements to the precise value of tilting angle $\chi$ of the goniometer [1]. Moreover, this is the reason why the standard texture measurement – pole figure – cannot reveal this fine feature of the layer structure.

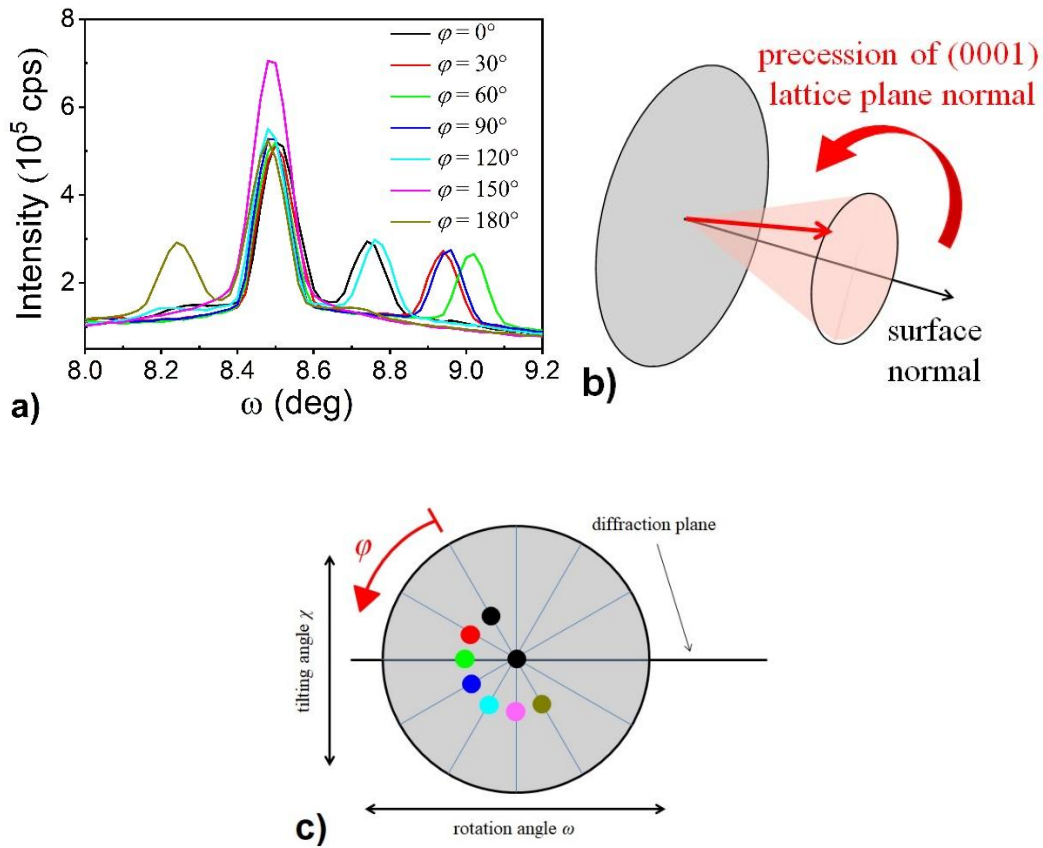

**Figure S2.** (a) $\omega$-scans of the diffraction 0001 of PtSe$_2$ layer measured at various azimuthal orientation $\varphi$. (b) Perspective view of the sample surface normal vector and precessing diffraction vector. (c) View in the direction against the surface normal. The coloured circles mark the position of diffraction vectors of PtSe$_2$ with respect to the diffraction plane and the surface normal. The directions of the rotation $\omega$ and tilting $\chi$ are also shown.

Further analysis revealed that the declination of PtSe$_2$ domains can be associated with the miscut of the sapphire substrates. In practice, the single crystalline substrates are never cut precisely parallel with the particular lattice planes. A slight miscut up to a few tenths of degree is admissible and has negligible effect on the process of layer growth. In order to reveal the possible role of the substrate miscut, the direction of the miscut was precisely measured. In addition to the $\omega$-scan of the 0001 diffraction of PtSe$_2$, $\omega$-scan of the diffraction 0006 of the sapphire substrate was measured for four azimuthal angles $\varphi$. The results are given in Fig. S3a, b. The positions of diffraction vectors 0006 of sapphire and 0001 of PtSe$_2$ are shown in Fig. S3c as coloured full and open circles, respectively. Evidently, the position of moving maximum in PtSe$_2$ 0001 $\omega$-scans correlates with the direction of substrate miscut. Based on these findings one can conclude that the PtSe$_2$ layers comprise of two slightly misoriented "sublayers" – one perfectly matched to the surface plane and one declined in the direction of substrate miscut.

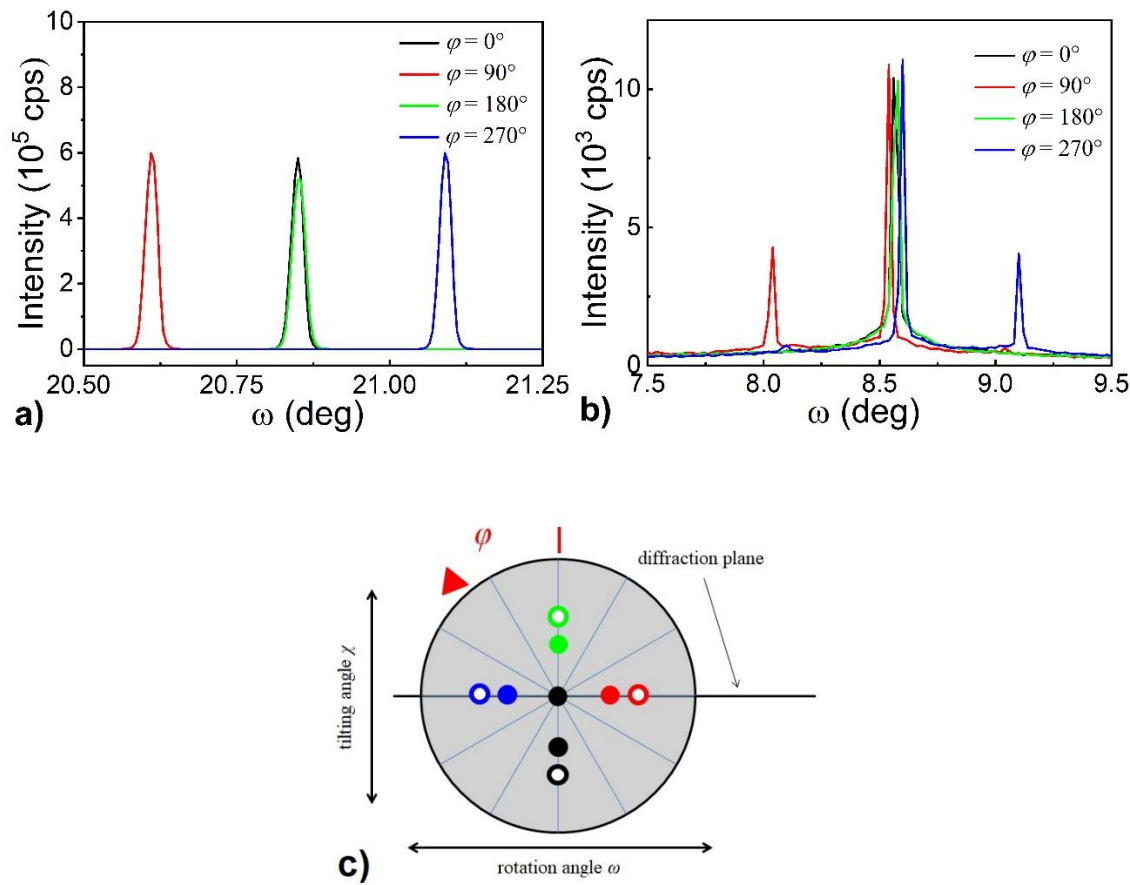

**Figure S3.** (a) $\omega$-scans of the diffraction 0006 of sapphire substrate measured at four azimuthal orientation $\varphi$. (b) $\omega$-scans of the diffraction 0001 of PtSe$_2$ layer measured at four azimuthal orientation $\varphi$. (c) View in the direction against the surface normal. The coloured full and open circles mark the position of diffraction vectors of sapphire and PtSe$_2$, respectively, with respect to the diffraction plane and the surface normal. The direction of the rotation $\omega$ and tilting $\chi$ are also shown.

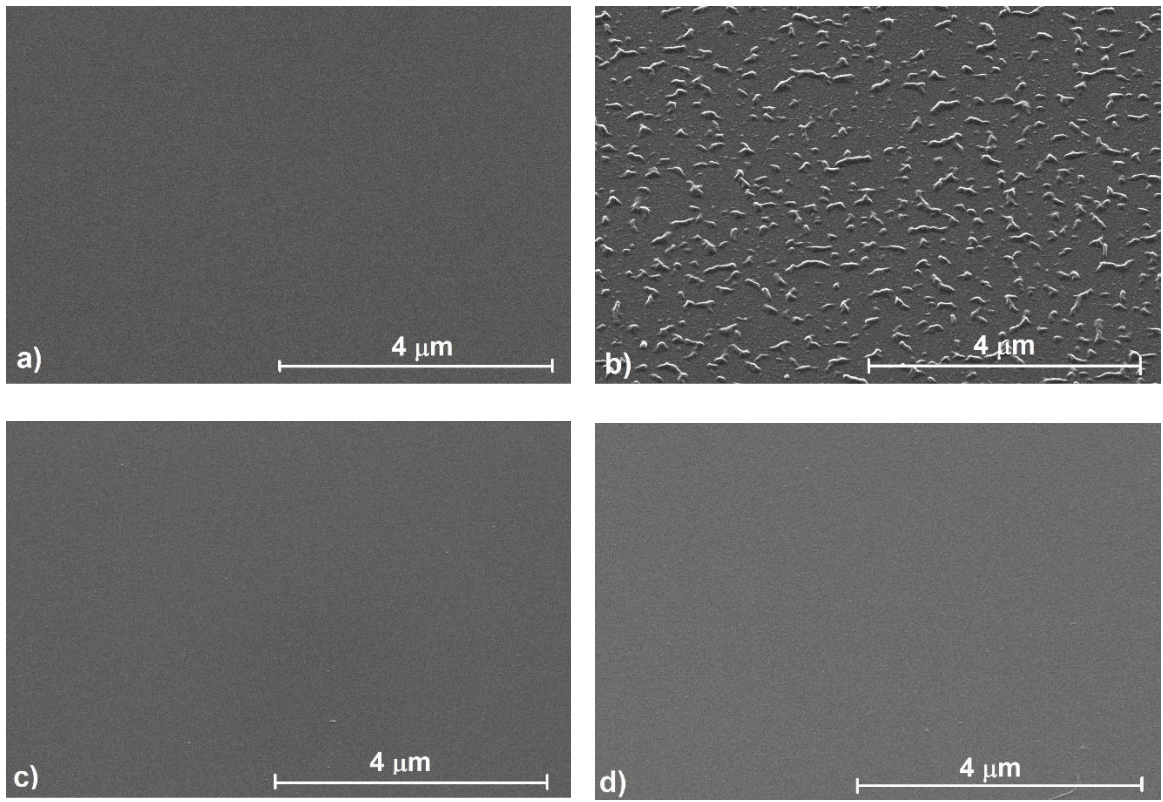

**Figure S4.** SEM image of PtSe$_2$ films prepared at 400 °C / 30 min from (a) 1 nm and (b) 3 nm thick Pt film and of PtSe$_2$ films made at 400 °C / 120 min from (c) 1 nm and (d) 3 nm thick Pt film.

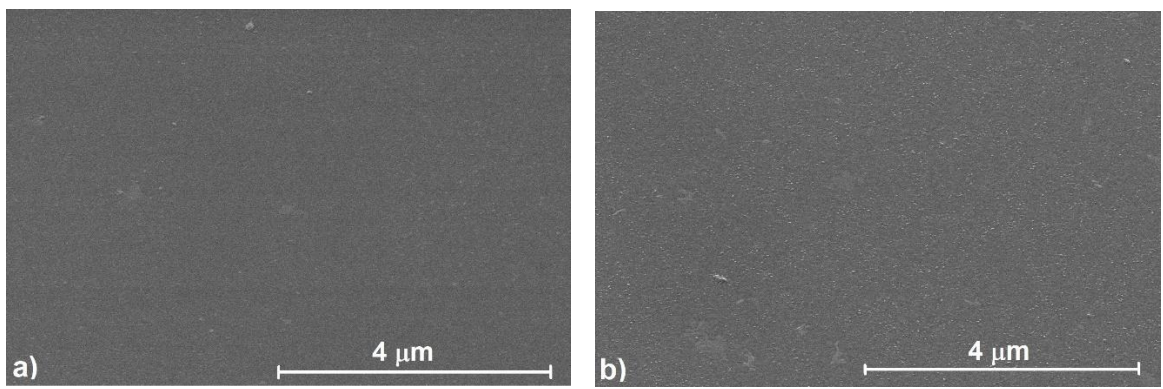

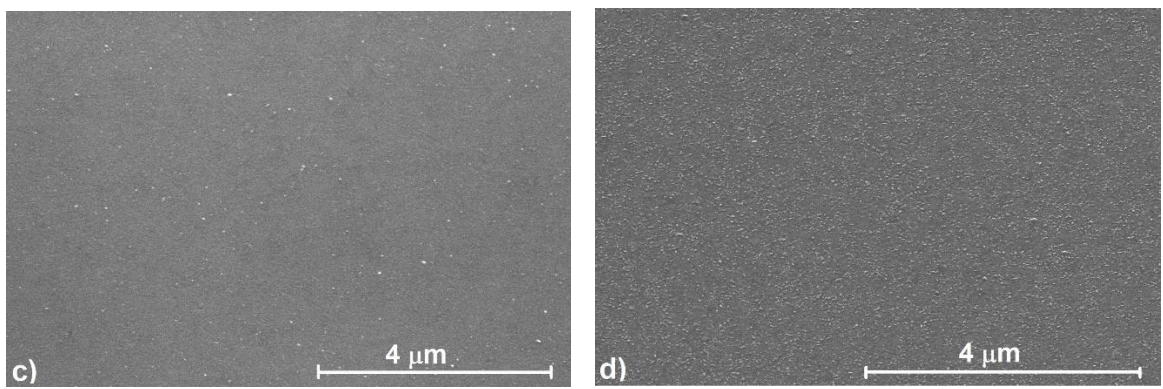

**Figure S5.** SEM image of PtSe$_2$ films prepared at 600 °C / 30 min from (a) 1 nm and (b) 3 nm thick Pt film and of PtSe$_2$ films fabricated at 600 °C / 120 min from (c) 1 nm and (d) 3 nm thick Pt film.

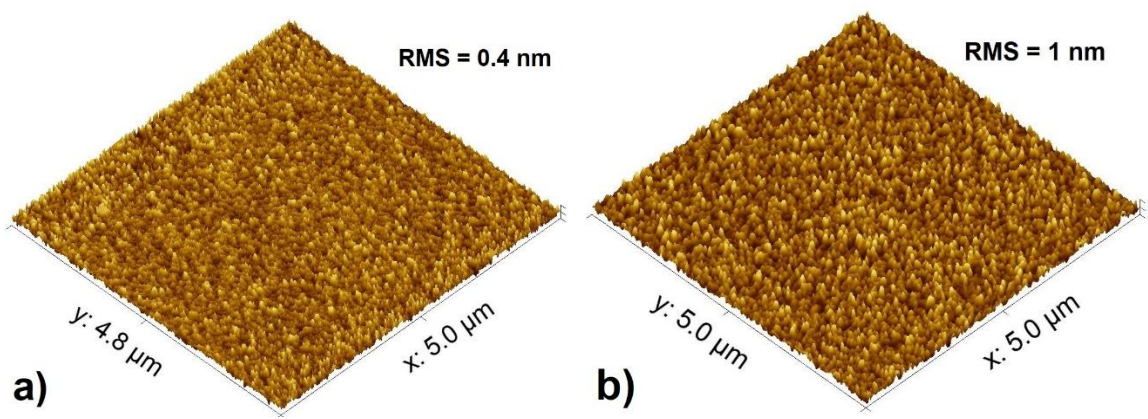

**Figure S6.** AFM image of PtSe$_2$ films prepared from 3 nm thick Pt film at (a) 400 °C / 120 min and (a) 600 °C / 120 min.

**References**

[1] M. Španková, E. Dobročka, V. Štrbík, Š. Chromik, N. Gál, N. Nedelko, A. Ślawska-Waniewska, P. Gierłowski, Structural Characterization of Epitaxial LSMO Thin Films Grown on LSAT Substrates, Acta Physica Polonica A. 137 (2020) 744–746. https://doi.org/10.12693/APhysPolA.137.744.